\documentclass[prl,twocolumn,showpacs,preprintnumbers,amsmath,amssymb]{revtex4}
\usepackage{graphicx}
\usepackage{dcolumn}
\usepackage{psfrag}
\usepackage{multirow}

\newcolumntype{e}[1]{D{,}{}{#1}}
\newcolumntype{f}[1]{D{.}{.}{#1}}

\DeclareMathOperator{\sTrace}{sTr}
\DeclareMathOperator{\diag}{diag}

\newcommand{\ms}{m_S}
\newcommand{\mv}{m_V}
\newcommand{\comma}{\quad,}
\newcommand{\sTr}[1]{\sTrace\bigl[#1\bigr]}
\newcommand{\pad}[1]{~#1~}

\begin{document}

\preprint{hep-lat/0112029}

\title{Is strong $CP$ invariance due to a massless up quark?}
\author{Daniel R. Nelson}
\author{George T. Fleming}
\author{Gregory W. Kilcup}
\affiliation{Department of Physics, The Ohio State University, Columbus, OH,
43210, USA}

\date{\today}

\begin{abstract}
A standing mystery in the Standard Model is the unnatural smallness of the
strong $CP$ violating phase.  A massless up quark has long been proposed as one
potential solution.  A lattice calculation of the constants of the chiral
Lagrangian essential for the determination of the up quark mass,
$2\alpha_8-\alpha_5$, is presented.  We find $2\alpha_8-\alpha_5 = 0.29 \pm
0.18$, which corresponds to $m_u / m_d = 0.410 \pm 0.036$.  This is the first
such calculation using a physical number of dynamical light quarks, $N_f = 3$.
\end{abstract}


\maketitle

\section{Introduction}

The nontrivial topological structure of the QCD gauge vacuum generates a $CP$
breaking term in the QCD Lagrangian.  However, measurements of the neutron
electric dipole moment have placed a restrictive upper bound on this term's
coefficient, $\bar{\theta} \le 10^{-9}$ \cite{PDBook}.  The unnatural smallness
of $\bar{\theta}$ is known as the strong $CP$ problem.

A massless up quark ($m_u = 0$) has long been proposed as a potential elegant
solution to the problem.  Chiral rotations of the quark mass matrix
$\mathcal{M}$ shift $\bar{\theta}$,
\begin{equation}
\bar{\theta} = \theta + \arg \det \mathcal{M} \comma
\end{equation}
where $\theta$ is a fundamental parameter of the Standard Model.  However, if
$m_u = 0$, then $\det \mathcal{M} = 0$ and $\arg \det \mathcal{M}$ is
unphysical, leaving one free to remove the $CP$ violating term through a simple
field redefinition.  

At leading order, Chiral Perturbation Theory (ChPT) appears to rule out the
possibility of $m_u = 0$.  The quark mass ratios, including $m_u / m_d$, can be determined using ChPT's LO predictions for the light meson masses.

At NLO, however, new coefficients appear in the chiral expansion which
contribute to the meson masses.  The parameters of ChPT are no longer fully
determined by experimental data.  In fact, it is impossible for ChPT to
distinguish between the effects of a non-zero up quark mass and certain large
NLO corrections.  This is known as the Kaplan-Manohar ambiguity
\cite{Kaplan:1986ru}.

Distinguishing between a light and a massless up quark requires knowledge of
the coefficients of the NLO terms in the chiral Lagrangian, the
Gasser-Leutwyler (GL) coefficients.  Specifically, it is the combination of
constants $2\alpha_8 - \alpha_5$ \footnote{The constants $\alpha_i$
are related to the corresponding constants of the chiral Lagrangian $L_i$ by
$\alpha_i = 8(4\pi)^2L_i$.} which appears in $\Delta_M$, the NLO correction to
the quark mass ratios \cite{Gasser:1985gg}.  If this combination falls within a
certain range, $-3.3 < 2\alpha_8 - \alpha_5 < -1.5$, current experimental
results can not rule out $m_u = 0$.

Various assumptions and phenomenological arguments have been used in the past
to assemble a somewhat standard set of values for the coefficients
\cite{Bijnens:1994qh}.  However, because these coefficients are physically
determined by the low-energy non-perturbative behavior of QCD, the lattice
offers the best opportunity for a truly first-principles calculation.

\section{Partially Quenched Chiral Perturbation Theory (\lowercase{pq}C\lowercase{h}PT)}

pqChPT \cite{Bernard:1992mk,Bernard:1994sv} is the tool through which one can
calculate the GL coefficients on the lattice.  pqChPT is distinct from standard
ChPT in that it is constructed from the symmetry of a graded group.  This
graded group follows from the presumed quark content of partially quenched QCD
(pqQCD):  separate valence and sea quark flavors in addition to ghost quark
flavors, which in perturbation theory cancel loop corrections due to valence
quarks.

The Lagrangian of pqChPT up to $O(p^4)$ follows, with only relevant NLO terms
shown.
\begin{align}
\mathcal{L} \pad{=} & \frac{f^2}{4} \sTr{ \partial_\mu U \partial^\mu U^\dagger
} - \frac{f^2}{4} \sTr{ \chi U^\dagger + U \chi } \nonumber \\
& + L_4 \sTr{ \partial_\mu U \partial^\mu U^\dagger } \sTr { \chi U^\dagger + U
\chi } \nonumber \\
& + L_5 \sTr{ \partial_\mu U \partial^\mu U^\dagger \left( \chi U^\dagger + U
\chi \right) } \nonumber \\
& - L_6 \sTr{ \chi U^\dagger + U \chi } \sTr{ \chi U^\dagger + U \chi }
\nonumber \\
& - L_8 \sTr{ \chi U^\dagger \chi U^\dagger + U \chi U \chi } + \: \cdots \comma
\end{align}
where $U = \exp \left( 2 i \Phi / f \right)$, $\chi = 2 \mu a^{-1} \diag \left(
\left\{ \ms, \mv \right\} \right)$, $\Phi$ contains the pseudo-Goldstone
``mesons'' of the spontaneously broken $SU(N_f + N_V | N_V)_L \otimes
\linebreak[0] SU(N_f + N_V | N_V)_R$ symmetry, and $U$ is an element of that
group.  $\ms$ and $\mv$ refer to the bare lattice quark mass parameters, which
are related to their dimensionful equivalents via the lattice spacing,
$m^{\text{dim}}_x = a^{-1} m_x$.  Three degenerate sea quarks were used, $N_f =
3$, while the number of valence quarks $N_V$ cancels in all expressions,
affecting only the counting of external states.  The constants $f$, $\mu$, and
the $L_i$'s are unknown, determined by the low-energy dynamics of pqQCD.

\begin{table*}
\caption{\label{table:simulation_details} Simulation details.}
\begin{ruledtabular}
\begin{tabular}{ccf{1.4}f{1.3}ccc|e{4.4}e{4.6}|e{3.5}e{3.5}|e{1.7}e{1.8}}
$L$ &
$T$ &
\multicolumn{1}{c}{$\beta$} &
\multicolumn{1}{c}{$\ms$} &
start\footnote{Starting configuration state:  ordered, disordered, or thermal.} &
traj &
\multicolumn{1}{c}{block} &
\multicolumn{2}{c}{$a^{-1}$ (MeV)} &
\multicolumn{2}{c}{$M_\pi(\mv=\ms)$ (MeV)} &
\multicolumn{2}{c}{$2\alpha_8-\alpha_5$} \\
\hline
& & & & & & \multicolumn{1}{c}{} & & \multicolumn{1}{c}{hyp\footnote{Denotes a hypercubic blocked ensemble.}} & & \multicolumn{1}{c}{hyp\footnotemark[3]} & & \multicolumn{1}{c}{hyp\footnotemark[3]} \\
\hline \hline
16 & 32 & 5.3   & 0.01  & O & 250-2250 & 200 & 1271,(85)   & 1376,.9(74) & 378,(25) & 271,.3(19) & 0,.236(12) & 0,.287(18) \vspace{4mm} \\
   &    &       &       & D & 250-2250 & 200 & & & & \\
\hline
12\footnote{Spatial volume is $12^2 \times 16$.} & 32 & 5.3   & 0.01  & O & 250-1850 & 200 & 1470,(130)  & 1419,(26)   & 438,(39) & 289,.1(70) & 0,.196(15) & 0,.226(64) \\
\hline
8 & 32 & 5.115  & 0.015 & O & 300-10300 & 100 & 679,.8(14) & 710,.9(24)  & 214,.58(45) & 218,.00(75) & 0,.326(12) & 0,.3439(76) \\
\hline
8 & 32 & 5.1235 & 0.02  & O & 300-10300 & 100 & 683,.5(12) & 723,.3(22)  & 249,.27(44) & 254,.18(79) & 0,.343(11) & 0,.3817(87)  \\
\hline
8 & 32 & 5.132  & 0.025 & T & 0-10000   & 100 & 686,.1(15) & 734,.6(22)  & 279,.54(62) & 286,.79(88) & 0,.388(10) & 0,.4150(91) \\
\hline
8 & 32 & 5.151  & 0.035 & T & 0-10000   & 100 & 695,.0(14) & 744,.3(25)  & 334,.45(68) & 341,.0(12) & 0,.475(12) & 0,.4704(94) \\
\hline
16 & 32 & 5.8   & \multicolumn{1}{c}{$\infty$} & & \multicolumn{2}{c|}{144 configs} & & 1408,.4(42) & & 274,.2(13) & & 0,.231(31) \\
\end{tabular}
\end{ruledtabular}
\end{table*}

Because the valence and sea quark mass dependence of the Lagrangian of pqChPT
is explicit and full QCD is within the parameter space of pqQCD ($\mv = \ms$),
the values obtained for the GL coefficients in a pqQCD calculation are the
exact values for the coefficients in full QCD \cite{Sharpe:2000bc,
Cohen:1999kk}.  Furthermore, independent variation of valence and sea quark
masses allows additional lever arms in the determination the the
coefficients.  Because the $N_f$ dependence of the Lagrangian is not explicit,
the GL coefficients are functions of $N_f$.  Thus, it is important to use a
physical number of light sea quarks, as we have, when extracting physical
results.  

\begin{figure}
\begin{center}
\psfrag{mVmK}[l]{\huge $\mv / m_K$}
\psfrag{mpMeV}[l]{\huge $M_\pi^2$ (MeV$^2$)}
\includegraphics*[angle=0, width=0.465\textwidth]{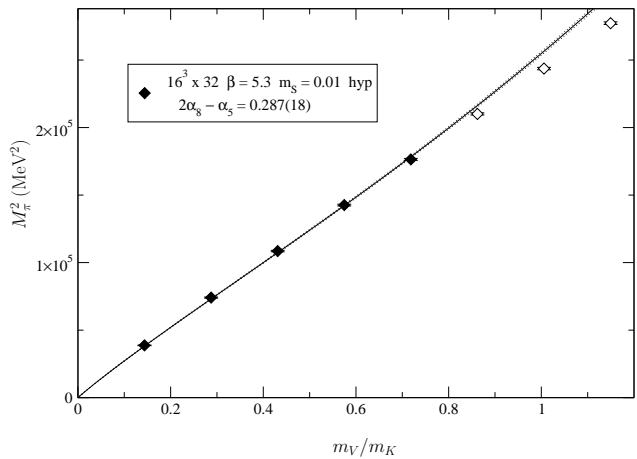}
\caption{$16^3 \times 32$, $\beta=5.3$, $\ms = 0.01$, hypercubic blocked.}
\label{figure:hyp_16_non}
\vspace{-8mm}
\end{center}
\end{figure}

\section{Predicted Forms}

pqChPT predicts forms for the dependence of the pseudoscalar mass and decay
constant on the valence quark mass, here assuming degenerate sea quarks and
degenerate valence quarks, and cutting off loops at $\Lambda_\chi = 4 \pi f$.
\begin{align}
M_\pi^2 \pad{=} & (4 \pi f)^2 z \mv \biggl\{ 1 \linebreak[0] + z \mv \Bigl( 2
\alpha_8 - \alpha_5 + \frac{1}{N_f} \Bigr) \biggr. \nonumber \\
& + \biggl. \frac{z}{N_f} \bigl( 2 \mv - \ms \bigr) \ln z \mv \biggr\}
\label{equation:m_pi_sqrd} \\
f_\pi \pad{=} & f \biggl\{ 1 + \frac{\alpha_5}{2} z \mv \biggr. \nonumber \\
& + \biggl. \frac{z N_f}{4} \bigl( \mv + \ms \bigr) \ln \frac{z}{2} \bigl( \mv
+ \ms \bigr) \biggr\} \label{equation:f_pi} \comma
\end{align}
where $z = 2 \mu a^{-1} / (4 \pi f)^2$.  These forms differ slightly from those
in \cite{Sharpe:1997by}, as the NLO dependence in the sea quark mass has been
absorbed into $\mu$ and $f$.  This is allowed as the error due to this change
is manifest when $z$ appears in the NLO terms, pushing the discrepancy up to
NNLO.  Accounting for these absorbed terms would require a systematic study at
several sea quark masses.

\begin{figure}
\begin{center}
\vspace{-8mm}
\psfrag{mVmK}[l]{\huge $\mv / m_K$}
\psfrag{RM}[l]{\huge $R_M$}
\includegraphics*[angle=0, width=0.465\textwidth]{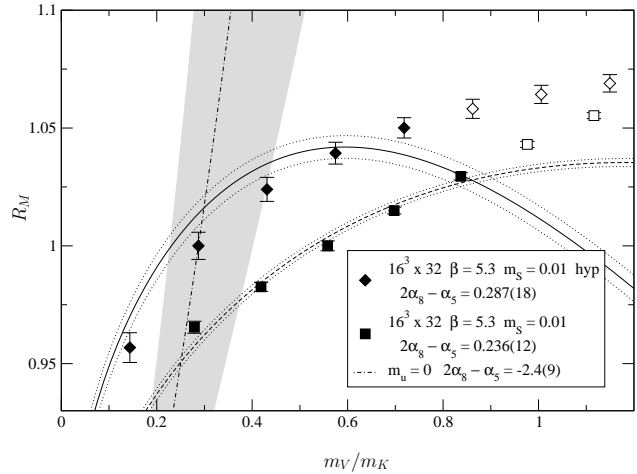}
\caption{$16^3 \times 32$, $\beta=5.3$, $\ms = 0.01$.  The
squares and dashed curve (diamonds and solid curve) are before (after)
hypercubic blocking.  Data consistent with $m_u = 0$ would fall within the
range centered on the dashed-dotted curve.}
\label{figure:hyp_16_zero}
\vspace{-8mm}
\end{center}
\end{figure}

The forms above are derived assuming degenerate light mesons.  However, our use
of staggered fermions on somewhat coarse lattices generates significant flavor
symmetry breaking, and thus a splitting of the light meson masses.  A detailed
analysis of this effect with dynamical quarks was recently presented in
\cite{Bernard:2001yj}.

In order to study this error, we applied hypercubic blocking to several of our
ensembles, using the blocking coefficients found in \cite{Hasenfratz:2001hp}.
Because hypercubic blocked dynamical quarks were not used when generating these
ensembles, we are using different Dirac operators for the valence and sea
quarks.  While this procedure may not have a clean continuum limit, it is still
useful for estimating the systematic error due to flavor symmetry breaking.

The ensembles were generated using staggered fermions via the inexact HMD
R-algorithm \cite{Gottlieb:1987mq}, which allows one to work at $N_f = 3$, but
involves taking the $\frac{3}{4}$ root of the quark determinant.  The result
is a quark action which is non-local at finite lattice spacing, but should
become local in the continuum limit.

To determine the value of $2\alpha_8-\alpha_5$, the local pseudoscalar
correlator was calculated using several valence quark masses.  The correlators
were fit to exponentials, while the meson mass and decay constants were
simultaneously fit to the predicted forms.  The results of the fit are the
values $\mu$, $f$, $\alpha_5$, and $2\alpha_8-\alpha_5$.  Figure
\ref{figure:hyp_16_non} displays an example of one such fit.

When fitting, a chiral cutoff point in the valence quark mass beyond which one
expects pqChPT to break down must be chosen.  To choose our cutoff, we added
$\mv$ values to our fit keeping $\mv / m_K \le 1$ and keeping $\chi^2 /
\mathrm{dof} \approx 1$.  The cutoff determined for the $16^3 \times 32$
hypercubic blocked ensemble, in terms of $\mv / m_K$, was used for all of the
$\beta = 5.3$, $\ms = 0.01$ ensembles.  $m_K$ denotes an ensemble's valence
quark mass at which the mass of the lightest pseudoscalar meson with degenerate
quarks equals the physical kaon mass.  We found $2\alpha_8-\alpha_5$ to be very
sensitive to our cutoff choice.  For the $16^3 \times 32$ hypercubic blocked
ensemble, changing the chiral cutoff by $\pm0.14$ in $\mv / m_K$ shifted
$2\alpha_8-\alpha_5$ by $\pm0.12$.

\begin{figure}
\begin{center}
\psfrag{mVmK}[l]{\huge $\mv / m_K$}
\psfrag{mSmK}[l]{\LARGE $\ms / m_K$}
\psfrag{RM}[l]{\huge $R_M$}
\psfrag{2a8a5}[l]{\LARGE $2\alpha_8-\alpha_5$}
\includegraphics*[angle=0, width=0.465\textwidth]{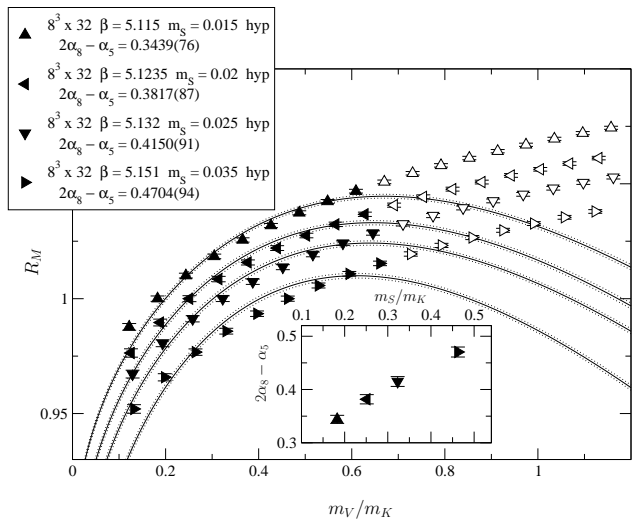}
\caption{$8^3 \times 32$, several values of $\beta$ and $\ms$ with
matched lattice spacing, hypercubic blocked.}
\label{figure:hyp_8x4}
\vspace{-8mm}
\end{center}
\end{figure}

Figures \ref{figure:hyp_16_zero}-\ref{figure:hyp_16_q} display the quantity
\begin{equation}
R_M = \frac{M^2_\pi(\ms)\mv}{M^2_\pi(\mv)\ms}
\end{equation}
suggested by \cite{Heitger:2000ay}.  This quantity accentuates the NLO terms in
$M^2_\pi$, as is evident by comparing Figure \ref{figure:hyp_16_non} to
Figures \ref{figure:hyp_16_zero}-\ref{figure:hyp_16_q}.  It should be
noted, however, that the full forms of $M^2_\pi$ and $f_\pi$,
\eqref{equation:m_pi_sqrd} and \eqref{equation:f_pi}, were used when fitting.
When calculating $R_M$, we did not use the simplification shown in
\cite{Heitger:2000ay}, but rather used a full numerator and denominator.

For each of the plots, the data points are the result of individual fits
of the correlator at each valence quark mass, with jackknife error bars.  The
curves display the result of a simultaneous fit of all the correlators below a
cutoff in $\mv / m_K$ to the predicted forms of $M_\pi^2$ and $f_\pi$,
\eqref{equation:m_pi_sqrd} and \eqref{equation:f_pi}, with jackknife
error bounds.  Solid symbols are used below the cutoff, while open symbols are
used beyond it.  Because of our small ensemble sizes, the full correlation
matrix for many of the ensembles proved to be nearly singular.  Thus, several
of the fits do not fully account for data correlation.  

We determined the lattice spacing of our ensembles via the static quark
potential, using a tree-level corrected Coulomb term.  The form of this term
for hypercubic blocked ensembles was taken from \cite{Hasenfratz:2001tw}.

\section{Results}

Figure \ref{figure:hyp_16_zero} presents $R_M$ for the $16^3 \times 32$
ensemble both before and after hypercubic blocking.  The application of
hypercubic blocking altered results significantly, suggesting that the effect
of flavor symmetry breaking at these lattice spacings is significant.  The
dashed-dotted curve uses the results from the hypercubic blocked ensemble's fit,
but replaces the value found for $2\alpha_8-\alpha_5$ with one consistent with a
zero up quark mass.  The data clearly fall well outside this range.

\begin{figure}
\begin{center}
\psfrag{mVmK}[l]{\huge $\mv / m_K$}
\psfrag{RM}[l]{\huge $R_M$}
\includegraphics*[angle=0, width=0.465\textwidth]{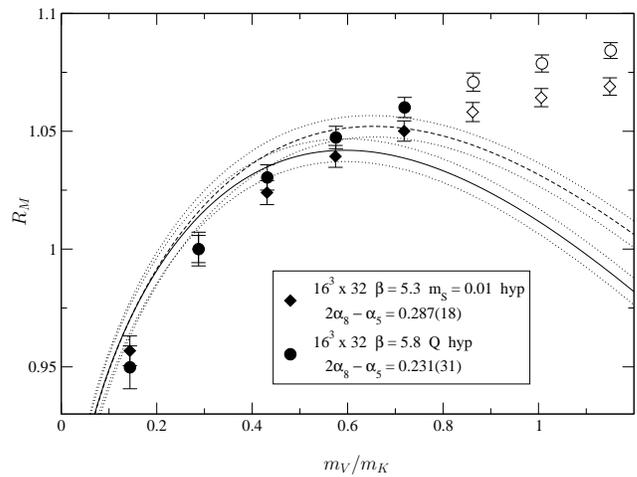}
\caption{$16^3 \times 32$, hypercubic blocked with similar lattice spacings.
The circles and dashed curve (diamonds and solid curve) are fully (partially)
quenched.  The quenched data were fit as though partially quenched with $m_s =
0.01$, $N_f = 3$.}
\label{figure:hyp_16_q}
\end{center}
\vspace{-8mm}
\end{figure}

To estimate the finite volume error in our result, we repeated the calculation
in a smaller $12^2 \times 16 \times 32$ volume, holding all other parameters
fixed.  Fitting this ensemble with the same chiral cutoff as the $16^3
\times 32$ ensemble resulted in the value $2\alpha_8-\alpha_5 = 0.226 \pm
0.064$.  This matches our quoted result, suggesting that the finite volume of
our $16^3 \times 32$ ensemble is not a large source of systematic error.

\begin{figure}
\begin{center}
\psfrag{2a8a5}[l]{\huge $2\alpha_8-\alpha_5$}
\includegraphics*[angle=0, width=0.465\textwidth]{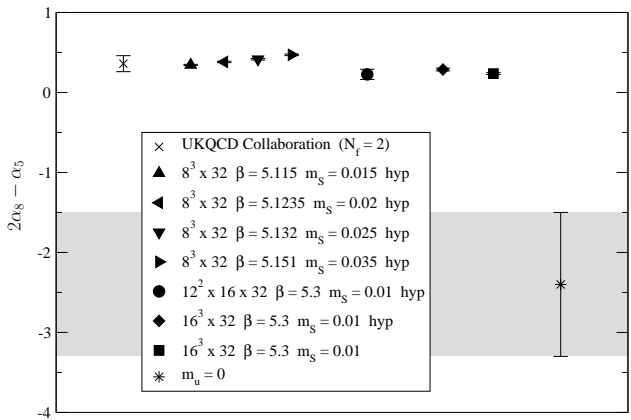}
\caption{Compiled results.  The UKQCD Collaboration data point is
taken from \cite{Irving:2001vy}, showing their statistical error only.  This
point was calculated using $N_f = 2$ Wilson fermions, and thus its relative
agreement suggests small $N_f$ dependence.  The rightmost point denotes the
range allowed by $m_u = 0$.}
\label{figure:results}
\end{center}
\vspace{-8mm}
\end{figure}

Figure \ref{figure:hyp_8x4} presents $R_M$ for several ensembles with a
variety of sea quark masses and matched lattice spacing.  The physical volume of
these lattices is similar to the physical volume of our $16^3 \times 32$
ensemble.  The Columbia group has determined several values of the critical
$\beta_c$ and $m_c$ for the $N_f = 3$, $N_t = 4$ finite temperature transition
\cite{Liao:2001en}.  These ensembles were generated using those bare
parameters.  The trend in $2\alpha_8-\alpha_5$ with the changing sea quark mass
is inset in Figure \ref{figure:hyp_8x4}.  This trend can be attributed to the
sea quark mass dependence which was dropped from Equations
\ref{equation:m_pi_sqrd} and \ref{equation:f_pi}.  A systematic study of this
dependence would allow a determination of the parameter within the dropped
term, $2\alpha_6-\alpha_4$ \cite{inprep}.

Fully quenched and partially quenched ChPT predict different forms for $R_M$.
Thus, one might hope to see the effects of quenching though an ensemble's $R_M$
plot.  Figure \ref{figure:hyp_16_q} shows the $16^3 \times 32$ partially
quenched hypercubic blocked ensemble along side a fully quenched ensemble with
similar lattice spacing.  The quenched ensemble was analyzed as though
partially quenched, using $\ms = 0.01$, $N_f = 3$.  This procedure does not
generate a rigorous value for $2\alpha_8-\alpha_5$, as this would require
the use of fully quenched ChPT.  However, it could offer insight into the
magnitude of quenching effects.  As Figure \ref{figure:hyp_16_q} shows, the
effects of quenching are not pronounced.  This analysis, at its current level
of precision, is unable to distinguish a fully quenched ensemble from a
partially quenched ensemble of equal lattice spacing.

The results for $2\alpha_8-\alpha_5$ from each ensemble are compiled in Figure
\ref{figure:results}.  While these values vary significantly, they do so well
outside the range required for a zero up quark mass, $-3.3 < 2\alpha_8-\alpha_5
< -1.5$.  Our quoted result of $2\alpha_8-\alpha_5 = 0.287 \pm
0.018^{\text{stat}} \pm 0.18^{\text{syst}}$ comes from our hypercubic blocked
$16^3 \times 32$ ensemble, where the reported systematic error is the result of
adding in quadrature the determined effects of shifting the chiral cutoff
$(\pm0.12)$, hypercubic blocking $(\pm0.05)$, doubling the lattice spacing
$(\pm0.11)$, and reducing the lattice volume $(\pm0.06)$.  Assuming Dashen's
rule \cite{Dashen:1969eg}, this corresponds to $\Delta_M = -0.0897 \pm 0.0313$
and $m_u / m_d = 0.484 \pm 0.027$, where the quoted error arises primarily from
the systematic error of our measurement.  The error from experimental input is
negligible and the size of NNLO corrections to $\Delta_M$ are assumed to be on
the order of $\Delta_M^2$.  Using the values for the electromagnetic
contributions to the light meson masses from \cite{Bijnens:1997kk} in place of
Dashen's rule results in $\Delta_M = -0.0898 \pm 0.0313$ and $m_u / m_d =
0.410 \pm 0.036$.  This can be compared to previous calculations in the
literature, which have given $m_u / m_d = 0.553 \pm 0.043$
\cite{Leutwyler:1996qg} and $m_u / m_d = 0.46 \pm 0.09$ \cite{Amoros:2001cp}.

\bibliography{letter1}

\end{document}